\PassOptionsToPackage{unicode}{hyperref}
\PassOptionsToPackage{hyphens}{url}
\documentclass[
]{article}
\usepackage{amsmath,amssymb}
\usepackage{lmodern}
\usepackage{ifxetex,ifluatex}
\ifnum 0\ifxetex 1\fi\ifluatex 1\fi=0 
  \usepackage[T1]{fontenc}
  \usepackage[utf8]{inputenc}
  \usepackage{textcomp} 
\else 
  \usepackage{unicode-math}
  \defaultfontfeatures{Scale=MatchLowercase}
  \defaultfontfeatures[\rmfamily]{Ligatures=TeX,Scale=1}
\fi
\IfFileExists{upquote.sty}{\usepackage{upquote}}{}
\IfFileExists{microtype.sty}{
  \usepackage[]{microtype}
  \UseMicrotypeSet[protrusion]{basicmath} 
}{}
\makeatletter
\@ifundefined{KOMAClassName}{
  \IfFileExists{parskip.sty}{%
    \usepackage{parskip}
  }{
    \setlength{\parindent}{0pt}
    \setlength{\parskip}{6pt plus 2pt minus 1pt}}
}{
  \KOMAoptions{parskip=half}}
\makeatother
\usepackage{xcolor}
\IfFileExists{xurl.sty}{\usepackage{xurl}}{} 
\IfFileExists{bookmark.sty}{\usepackage{bookmark}}{\usepackage{hyperref}}
\hypersetup{
  pdftitle={Interactive Dashboard to Monitor the COVID-19 Outbreak and Vaccine Administration},
  hidelinks,
  pdfcreator={LaTeX via pandoc}}
\usepackage[margin=1in]{geometry}
\usepackage{longtable,booktabs,array}
\usepackage{calc} 
\usepackage{etoolbox}
\makeatletter
\patchcmd\longtable{\par}{\if@noskipsec\mbox{}\fi\par}{}{}
\makeatother
\IfFileExists{footnotehyper.sty}{\usepackage{footnotehyper}}{\usepackage{footnote}}
\makesavenoteenv{longtable}
\usepackage{graphicx}
\makeatletter
\def\maxwidth{\ifdim\Gin@nat@width>\linewidth\linewidth\else\Gin@nat@width\fi}
\def\maxheight{\ifdim\Gin@nat@height>\textheight\textheight\else\Gin@nat@height\fi}
\makeatother
\setkeys{Gin}{width=\maxwidth,height=\maxheight,keepaspectratio}
\makeatletter
\def\fps@figure{htbp}
\makeatother
\providecommand{\tightlist}{%
  \setlength{\itemsep}{0pt}\setlength{\parskip}{0pt}}
\usepackage{url}
\usepackage{hyperref}
\ifluatex
  \usepackage{selnolig}  
\fi
\newlength{\cslhangindent}
\newlength{\csllabelwidth}
\newenvironment{CSLReferences}[2] 
 {
  \setlength{\parindent}{0pt}
  \ifodd #1 \everypar{\setlength{\hangindent}{\cslhangindent}}\ignorespaces\fi
  \ifnum #2 > 0
  \setlength{\parskip}{#2\baselineskip}
  \fi
 }%
 {}
\usepackage{calc}

\title{Interactive Dashboard to Monitor the COVID-19 Outbreak and
Vaccine Administration}
\author{}
\date{\vspace{-2.5em}}

\begin{document}
\maketitle

\textbf{Thiyanga S. Talagala}\footnote{Corresponding author, email:
  \href{mailto:ttalagala@sjp.ac.lk}{\nolinkurl{ttalagala@sjp.ac.lk}}}

Department of Statistics, Faculty of Applied Sciences

University of Sri Jayewardenepura, Sri Lanka, CO 10230

\hspace{3cm}

\textbf{Randi Shashikala}

Department of Statistics, Faculty of Applied Sciences

University of Sri Jayewardenepura, Sri Lanka, CO 10230

\hspace{3cm}

\textbf{Abstract}

Dashboards are the most common visualization method for displaying
COVID-19 data and informing the public. We examined 15 different
dashboards to see how various visualization techniques were used. This
paper describes the creation and implementation of a dashboard for
COVID-19 epidemic and vaccination administration data in Sri Lanka.


\hypertarget{introduction}{%
\section{Introduction}\label{introduction}}

COVID-19 has expanded over the globe, having a significant impact on our
daily lives and work. Early responses and timely decisions and actions
are critical to saving communities and economies worldwide. Data is
essential in order to make effective decisions. Data-driven information
guides the decision-making process and also evaluates the effectiveness
of strategies taken.

Massive amounts of data are being generated in the response to the
COVID-19 pandemic. Given this available data, it is critical to create
tools for exploratory analysis for policy-makers, health officials, and
the general public. Dashboards are one of the greatest visual
interpretation methods for tracking the COVID-19 pandemics spread and
vaccine administration. Dashboards allow users to quickly interact with
a combination of exploratory visualizations and gain a quick overview of
the data. This paper describes the development and implementation of a
dashboard for the COVID-19 outbreak and vaccine administration data in
Sri Lanka.

There are a plethora of COVID-19 visualization dashboards that have been
designed to visualize the pandemics global and local status. Different
software can be used to generate dashboards. We explored 15 dashboards
designed to visualize COVID-19 data at the global and country levels.
First, dashboards were compared to identify the various features,
visualization approaches, and enhancements that should be implemented.
Next, we developed an interactive dashboard to visualize the COVID-19
outbreak and vaccination information in Sri Lanka. This dashboard
provides front-line health officers a situational awareness of the
spread of COVID-19 and the status of the vaccination program.

The rest of the paper is organized as follows:
\protect\hyperlink{litreview}{Section 2} of dashboards created using
data related to the COVID-19 pandemic.
\protect\hyperlink{methods}{Section 3} presents the methodology and
basic design concept; \protect\hyperlink{results}{Section 4} presents
the results; and \protect\hyperlink{ux5cux2520conclusion}{Section 5}
concludes.

\hypertarget{litreview}{%
\section{Literature Review}\label{litreview}}

Dashboards are one of the best visual interpretation methods for
tracking the spread and communication of the COVID-19 pandemic. The 15
dashboards we used in the literature survey are listed in Table 1. We
compared dashboards to identify data types, plotting techniques, colour
themes, and other features such as interactivity on plots and panel
numbers.

\textbf{Table 01: Labels of the dashboards}

\begin{longtable}[]{@{}
  >{\centering\arraybackslash}p{(\columnwidth - 4\tabcolsep) * \real{0.13}}
  >{\raggedright\arraybackslash}p{(\columnwidth - 4\tabcolsep) * \real{0.67}}
  >{\raggedright\arraybackslash}p{(\columnwidth - 4\tabcolsep) * \real{0.20}}@{}}
\toprule
No & Name of the Dashboard & Reference \\
\midrule
\endhead
1 & COVID-19 dashboard created by the John Hopkins University Centre for
Systems Science \& Engineering (JHU CSSE) & {``{COVID-19 Map -- Johns
Hopkins Coronavirus Resource Center}''} (2022) \\
2 & WHO COVID-19 Dashboard & {``{WHO Coronavirus (COVID-19)
Dashboard}''} (2021) \\
3 & COVID-19 surveillance dashboard created by the University of
Virginia & {``{COVID-19 Surveillance Dashboard -- NSSAC Research}''}
(2022) \\
4 & Corona cases (COVID-19) per municipality in Belgium dashboard &
{``{COVID 19 Dashboard -- Belgium}''} (2022) \\
5 & COVID-19 dashboard for England created by NHS providers & {``{COVID
19 Dashboard -- NHS providers}''} (2022) \\
6 & NZ COVID-19 Dashboard & {``{New Zealand COVID-19 Surveillance
Dashboard}''} (2021) \\
7 & Pakistan's official COVID-19 dashboard & {``{Pakistan's Official
COVID-19 Dashboard -- Shifa International Hospitals Ltd}''} (2021) \\
8 & COVID-19 Canada live dashboard & {``{Track COVID-19 Across Canada
Using Our Interactive Dashboards}''} (2021) \\
9 & India (COVID-19) Dashboard & {``{COVID 19 Dashboard India -- ZOHO
Analytics -- ZOHO}''} (2022) \\
10 & Italy COVID-19 dashboard & {``{COVID-19 integrated surveillance
data in Italy -- EpiCentro}''} (2022) \\
11 & Jamaica COVID-19 Dashboard & {``{COVID-19 Jamaica - Ministry of
Health and Wellness}''} (2021) \\
12 & GCI COVID-19 dashboard for Russia & {``{The Global COVId-19 Index
(GCI) -- Russia Dashboard -- PEMANDU Associates}''} (2021) \\
13 & COVID-19 live situation analysis dashboard of Sri Lanka &
{``{COVID-19: Live situational Analysis Dashboard of Sri Lanka}''}
(2022) \\
14 & COVID 19 ZA South Africa Dashboard & {``{COVID-19 ZA Dashboard -
Data Studio}''} (2021) \\
15 & COVID-19 dashboard for Germany & {``{RKI COVID-19 Germany -- ArcGIS
Experience}''} (2021) \\
\bottomrule
\end{longtable}

Table 02 summarizes the data types that are most frequently shown in
dashboards. As shown in Table 02, all dashboards which are considered in
this paper represent the data related to COVID-19 confirmed cases,
recovered cases, and deaths. There were 8 dashboards out of 15
dashboards that contained vaccination details.

\textbf{Table 02: Summary of data represent in the dashboards}

\begin{longtable}[]{@{}
  >{\centering\arraybackslash}p{(\columnwidth - 14\tabcolsep) * \real{0.13}}
  >{\raggedright\arraybackslash}p{(\columnwidth - 14\tabcolsep) * \real{0.12}}
  >{\centering\arraybackslash}p{(\columnwidth - 14\tabcolsep) * \real{0.13}}
  >{\centering\arraybackslash}p{(\columnwidth - 14\tabcolsep) * \real{0.13}}
  >{\centering\arraybackslash}p{(\columnwidth - 14\tabcolsep) * \real{0.13}}
  >{\centering\arraybackslash}p{(\columnwidth - 14\tabcolsep) * \real{0.13}}
  >{\centering\arraybackslash}p{(\columnwidth - 14\tabcolsep) * \real{0.13}}
  >{\centering\arraybackslash}p{(\columnwidth - 14\tabcolsep) * \real{0.13}}@{}}
\toprule
Name of the Dashboard & Location (Represented) & Confirmed Cases &
Recovered Cases & Deaths & Vaccination Details & Tests & Global
Comparison \\
\midrule
\endhead
1 & Global & \checkmark & \checkmark & \checkmark & \checkmark & &
\checkmark \\
2 & Global & \checkmark & \checkmark & \checkmark & \checkmark & &
\checkmark \\
3 & Global & \checkmark & \checkmark & \checkmark & & \checkmark & \\
4 & Belgium & \checkmark & \checkmark & \checkmark & & & \\
5 & England & \checkmark & \checkmark & \checkmark & \checkmark & & \\
6 & New Zealand & \checkmark & \checkmark & \checkmark & & &
\checkmark \\
7 & Pakistan & \checkmark & \checkmark & \checkmark & & \checkmark & \\
8 & Canada & \checkmark & \checkmark & \checkmark & & & \\
9 & India & \checkmark & \checkmark & \checkmark & \checkmark & &
\checkmark \\
10 & Italy & \checkmark & \checkmark & \checkmark & & \checkmark & \\
11 & Jamaica & \checkmark & \checkmark & \checkmark & \checkmark & & \\
12 & Russia & \checkmark & \checkmark & \checkmark & & & \\
13 & Sri Lanka & \checkmark & \checkmark & \checkmark & & \checkmark &
\checkmark \\
14 & South Africa & \checkmark & \checkmark & \checkmark & \checkmark &
\checkmark & \\
15 & German & \checkmark & \checkmark & \checkmark & \checkmark & & \\
\bottomrule
\end{longtable}

Table 03 highlights the dashboard visualization techniques. Value boxes
have been utilized to display total figures on practically every
dashboard. The most common ways of visualizing confirmed cases,
recovered cases, deaths, and immunization details are bar charts and
line charts (trend lines). The majority of dashboards displayed data on
a daily or weekly basis. The spatial distribution of COVID-19 cases by
country, province, regional, and other factors is tracked using
choropleth maps. When visualizing the data by the map colour code
system, circles with respect to the size of the cases have been used to
visualize the variation in size. Several dashboards use doughnut-shaped
pie charts to indicate total COVID-19 confirmed cases, recovered cases,
active cases, and deaths as a proportion. Furthermore, region, gender,
age group, and ethnicity can be identified as common breakdowns of
COVID-19 cases. Data tables for representing cases' distribution by
province/region have been added to some dashboards. Very few dashboards
have been visualized in the COVID-19 test details. Only 6 dashboards
have been compared to global situations. In addition, the fatality rate,
incidence rate, ICU beds, stage of the patients, and hospitalized
details have been contained in the several dashboards.\hfill\break

\textbf{Table 03: Summary of tools which are used for different purpose}

\begin{longtable}[]{@{}lccccccc@{}}
\toprule
Purpose & Bar chart & Line chart & Pie chart & Dot plot & Heat map &
Mapping & Data table \\
\midrule
\endhead
COVID-19 confirmed & & & & & & & \\
cases & \checkmark & \checkmark & & \checkmark & & \checkmark &
\checkmark \\
COVID-19 deaths & \checkmark & \checkmark & & & & \checkmark &
\checkmark \\
COVID-19 recovered & & & & & & & \\
cases & \checkmark & \checkmark & & & & \checkmark & \checkmark \\
COVID-19 vaccination & & \checkmark & & & & \checkmark & \checkmark \\
COVID-19 test & & & & & & & \\
conducted & \checkmark & \checkmark & & & & & \\
Clinical status & \checkmark & & & & & & \\
Cases distribution by & & & & & & & \\
age & \checkmark & & \checkmark & & & & \\
Cases distribution by & & & & & & & \\
gender & \checkmark & & & & & & \\
Cases distribution by area & & & & & & & \\
(Province/state/region) & \checkmark & \checkmark & & & \checkmark &
\checkmark & \checkmark \\
To compare the cases & & & \checkmark & & & & \checkmark \\
Global comparison & \checkmark & \checkmark & & & & \checkmark &
\checkmark \\
\bottomrule
\end{longtable}

\hypertarget{comparison-of-dashboards}{%
\subsection{Comparison of Dashboards}\label{comparison-of-dashboards}}

Before developing a dashboard, it is necessary to think about which
visualization tools and features should be contained in the dashboard.
What are the most suitable plots, how many panels in the dashboard, what
data should be included, how to fit the dashboard on a screen, colours,
and is it real time updated or not are the common things that should be
considered before developing the dashboards. Table 4 summarizes
information under the following categories:

\begin{enumerate}
\def\labelenumi{\roman{enumi}.}
\tightlist
\item
  Number of panels - How many panels are included in the dashboard.
\item
  Visualization tools -- what are the graphical representations of data.
\item
  Fitted on a single screen -- whether the dashboard fits on a single
  screen or not (users can see the whole dashboard on a single screen
  without adjusting through grid overlay or not).
\item
  Colour theme -- is there a unique colour used for one data type in the
  whole dashboard (i.e.: one colour scale for one data type everywhere
  on the dashboard).
\item
  Dark background -- the background colour of the dashboard is dark or
  light.
\item
  Data available -- whether users can downloaded or whether data is
  available to reproduce the results.
\item
  Real time updated -- whether the dashboard is updated daily/ specific
  time (live dashboard) or not.
\end{enumerate}

\textbf{Table 04: Comparison of visualization tools and features of
dashboard}

\begin{longtable}[]{@{}
  >{\centering\arraybackslash}p{(\columnwidth - 14\tabcolsep) * \real{0.12}}
  >{\centering\arraybackslash}p{(\columnwidth - 14\tabcolsep) * \real{0.12}}
  >{\centering\arraybackslash}p{(\columnwidth - 14\tabcolsep) * \real{0.12}}
  >{\centering\arraybackslash}p{(\columnwidth - 14\tabcolsep) * \real{0.12}}
  >{\centering\arraybackslash}p{(\columnwidth - 14\tabcolsep) * \real{0.12}}
  >{\centering\arraybackslash}p{(\columnwidth - 14\tabcolsep) * \real{0.12}}
  >{\centering\arraybackslash}p{(\columnwidth - 14\tabcolsep) * \real{0.12}}
  >{\centering\arraybackslash}p{(\columnwidth - 14\tabcolsep) * \real{0.12}}@{}}
\toprule
Name of the Dashboard & Number of panels & Visualization tools & Fitted
on a single screen & Colour theme & Dark background & Data available &
Real time updated \\
\midrule
\endhead
1 & 1 & Bar chart\hfill\break  Interactive map\hfill\break & \checkmark
& \checkmark & \checkmark & \checkmark & \checkmark \\
2 & 4 & Line chart\hfill\break  Interactive map\hfill\break  Data
table\hfill\break & & \checkmark & & \checkmark & \checkmark \\
3 & 2 & Line chart\hfill\break  Bar chart\hfill\break Interactive
map\hfill\break Data table\hfill\break & \checkmark & \checkmark &
\checkmark & \checkmark & \checkmark \\
4 & 1 & Line chart\hfill\break Bar chart\hfill\break  Pie
chart\hfill\break Interactive map\hfill\break & \checkmark & &
\checkmark & \checkmark & \checkmark \\
5 & 1 & Line chart\hfill\break Bar chart\hfill\break Data
table\hfill\break & & \checkmark & & \checkmark & \checkmark \\
6 & 5 & Line chart\hfill\break Bar chart\hfill\break Dot
plot\hfill\break  Interactive country map\hfill\break & & & & \checkmark
& \checkmark \\
7 & 1 & Line chart\hfill\break Bar chart\hfill\break  Country
map\hfill\break Data table\hfill\break & & \checkmark & & &
\checkmark \\
8 & 3 & Line chart\hfill\break Bar chart\hfill\break Data
table\hfill\break Interactive map\hfill\break & & & \checkmark & &
\checkmark \\
9 & 3 & Line chart\hfill\break Bar chart\hfill\break Doughnut shape pie
chart\hfill\break Data table\hfill\break Interactive country
map\hfill\break & & \checkmark & & \checkmark & \checkmark \\
10 & 2 & Bar chart\hfill\break Doughnut shape pie chart\hfill\break Heat
map\hfill\break Interactive country map\hfill\break & & & & &
\checkmark \\
11 & 1 & Line chart\hfill\break Bar chart\hfill\break Doughnut shape pie
chart\hfill\break Data table\hfill\break Interactive country
map\hfill\break & & \checkmark & & & \checkmark \\
12 & 1 & Line chart\hfill\break Bar chart\hfill\break  Interactive
map\hfill\break & & \checkmark & \checkmark & & \checkmark \\
13 & 1 & Line chart\hfill\break Bar chart\hfill\break Doughnut shape pie
chart\hfill\break & & \checkmark & & & \checkmark \\
14 & 2 & Line chart\hfill\break Bar chart\hfill\break  Interactive
Country map\hfill\break & & \checkmark & & \checkmark & \checkmark \\
15 & 1 & Line chart\hfill\break Bar chart\hfill\break  Data
table\hfill\break Interactive map\hfill\break & \checkmark & \checkmark
& \checkmark & & \checkmark \\
\bottomrule
\end{longtable}

As listed in Table 4, on almost each and every dashboard, line charts
and bar charts have been used to visualize the data. The heat map and
dot plot were only used on one dashboard. Only four dashboards have been
fitted with a single screen. The majority of dashboards use a colour
theme on the whole dashboard. That is, dashboards have been applied with
different colours for different types of data (i.e., one specific colour
for confirmed cases, another colour for deaths, etc.) across the whole
dashboard. The data set and related links are available on some
dashboards, and users can download these data sets. There are 6
dashboards with a dark background, while others have a light background.
The last updated time and date of the latest available data has been
reported at the top or bottom of the first panel in the dashboard. Half
of the dashboards included all the information in a single panel.

\hypertarget{method}{%
\section{Methodology}\label{method}}

\hypertarget{data}{%
\subsection{Data}\label{data}}

We obtained data from COVID-19 situation reports published by the
Epidemiology Unit, Ministry of Health Sri Lanka. The data includes the
number of death cases, number of hospitalized cases, number of recovered
cases, and COVID-19 vaccinated counts in Sri Lanka. The data is made
available through an open-source R package covid19srilanka (Talagala
2021).

\hypertarget{design-and-development}{%
\subsection{Design and development}\label{design-and-development}}

R software was used for data cleaning and analysis. The flexdashboard
(Iannone, Allaire, and Borges 2020) package was used to build the data
visualization dashboard. The initial layout for the dashboard was
prepared based on Krispin (2021). Data visualizations are generated
using the ggplot2 (\textbf{ggplot2?}) and plotly (Sievert 2020) packages
in R. We used colour-blind friendly colour palettes for the graphics. A
diverging colour palette was used to represent qualitative data, and to
represent numeric variables, a sequential colour theme was used. Table 5
provides an overview of methods that have been used to visualize data.

\textbf{Table 5: Data visualization approaches used to visualize data}

\begin{longtable}[]{@{}
  >{\raggedright\arraybackslash}p{(\columnwidth - 2\tabcolsep) * \real{0.54}}
  >{\raggedright\arraybackslash}p{(\columnwidth - 2\tabcolsep) * \real{0.46}}@{}}
\toprule
\textbf{Data} & \textbf{Type of graphics} \\
\midrule
\endhead
Daily COVID-19 confirmed & Time series plots \hfill\break \\
Daily COVID-19 recovered cases by time & Time series plots
\hfill\break \\
Daily COVID-19 death cases by time & Time series plots \hfill\break \\
Total COVID-19 confirmed cases by time and wave & Time series plots
annotated with vertical lines to denote significant milestones
\hfill\break \\
Total COVID-19 death cases by time and wave & Histogram \hfill\break \\
Distribution of COVID-19 patients by districts & Tree map, Choropleth
maps, Dorling Cartogram, heat maps \hfill\break \\
Total vaccination by first dose and second dose & Time series plot
\hfill\break \\
Total administrated does by vaccine name & Stacked bar chart
\hfill\break \\
Total administrated does by vaccine name & Stacked bar chart
\hfill\break \\
Total administrated does by vaccine name & Stacked bar chart
\hfill\break \\
Comparison of cases with in Sri Lanka with Top 10 countries & Cumulative
cases by time, Log of cumulative cases by time, stacked bar chart
\hfill\break \\
Spread of COVID-19 around the world & Choropleth maps \hfill\break \\
\bottomrule
\end{longtable}

We now describe the novel visualization approaches we included in our
dashboard. To effectively distribute the vaccine and to support
situational awareness and inform policy-makers' decision-making, it is
important to know the district-wise spread of COVID-19 cases. ~We have
daily COVID-19 data related to confirmed cases in all 25 districts in
Sri Lanka. This structure generates a multiple time series collection.
Visualizing this time series data is useful to identify similarities and
dissimilarities between districts and their general trends. There are
two approaches to visualizing these time series: (i) creating individual
time series plots for each district (as shown in Figure 1-A), and (ii)
plotting all time series on a single panel at the same time (as shown in
Figure 1-B). Plotting all time series simultaneously is also not
possible due to overlapping time series and scale differences. Plotting
separate panels for each district is not effective. The reason is that
it is hard to compare across 25 different panels at once. In order to
overcome these problems in multiple time series visualization, we use
heat maps (Peng (2008)) to visualize global and local similarities and
dissimilarities across districts. The associated results are shown in
Figure 2. Here, two heat maps are used to show the global variations
(Figure 2: A) and local variations (Figure 2: B) in the time series
collection. Figure 2-A cell colours represent the actual counts of the
COVID-19 confirmed cases. This is useful to get an idea of the
differences in absolute values. Figure 2-B cell colours represent the
normalized values created by applying the min-max transformation. A
min-max transformation is applied to each district's time series by
using the corresponding district's minimum and maximum value. This helps
us to get an idea about patterns within districts. For example,
according to Figure 2A, we can see that in Colombo, Gampaha, and
Kalutara districts, COVID-19 cases are significantly higher than in
other districts. According to Figure 2B, all districts show an
increasing trend pattern as the right-hand side of the cells are lighter
than the left-hand side cells in the heat map. Furthermore, according to
Figure 2B, all districts reported a high number of cases on August 19,
24, and 29, 2021. Figure 3B is useful for identifying these local
outlying behaviours. As shown in Figure 4, we also use a Choropleth map
and Dorling cartogram to visualize the spatial distribution of COVID-19
cases. The vaccination information is visualized through interactive
time series plots and bar charts.

\begin{figure}
\centering
\includegraphics{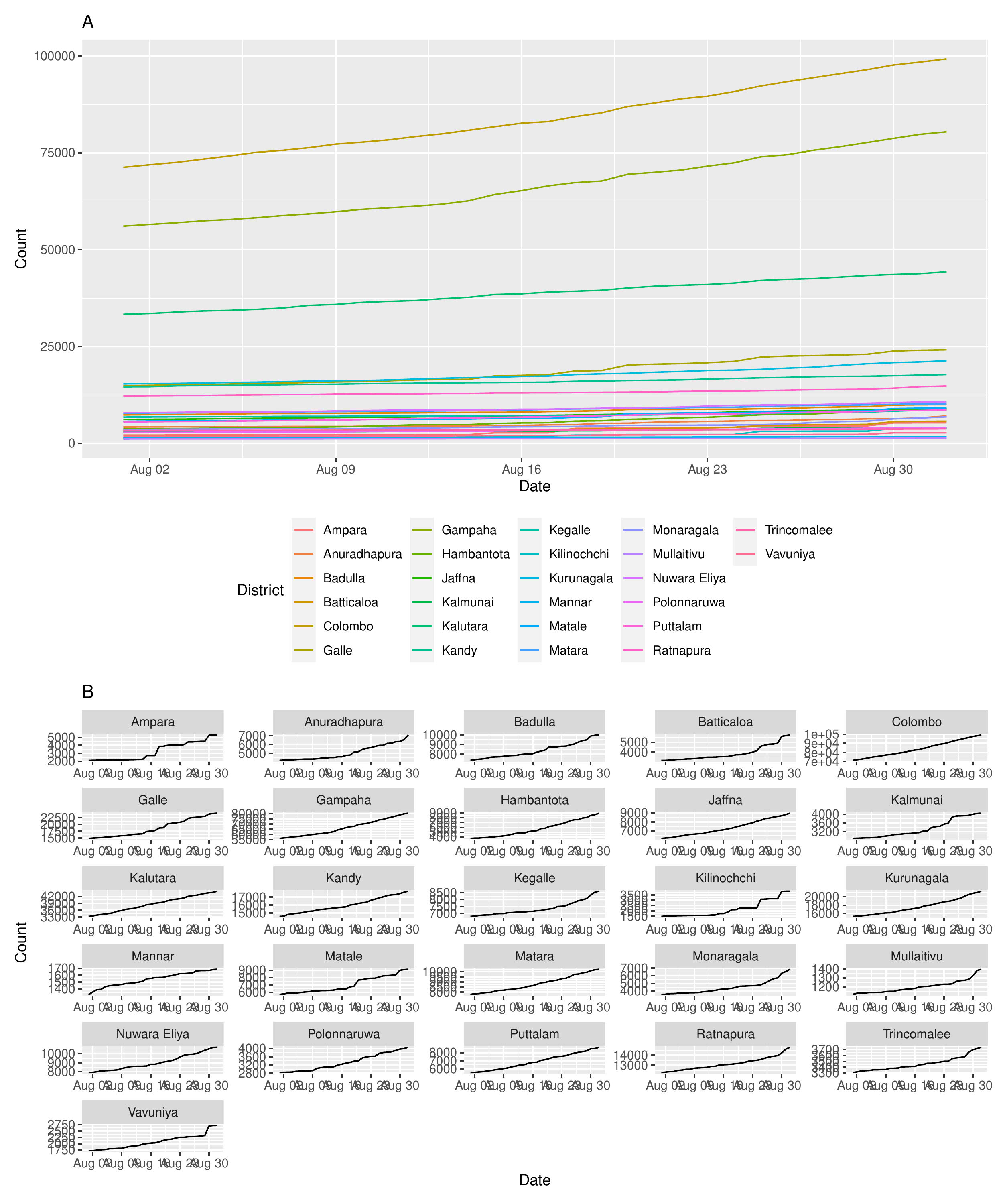}
\caption{Distribution of COVID-19 cases by districts: (A) Plotted on a
single panel, (B) Plotted on separate panels.}
\end{figure}

\begin{figure}
\centering
\includegraphics{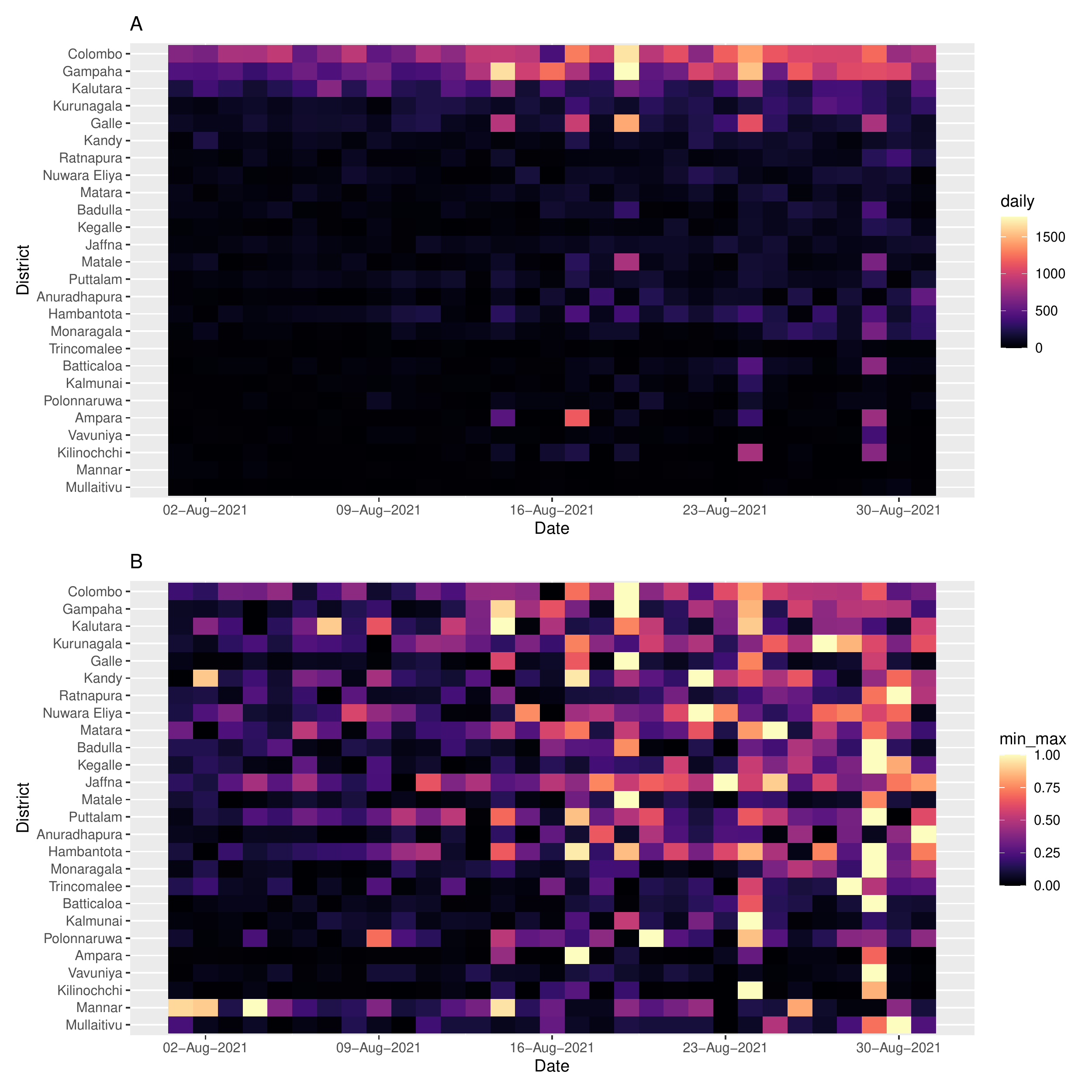}
\caption{(A) Global view of distribution of COVID-19 cases by districts,
(B) Local view of distribution of COVID-19 cases by districts}
\end{figure}

\begin{figure}
\centering
\includegraphics{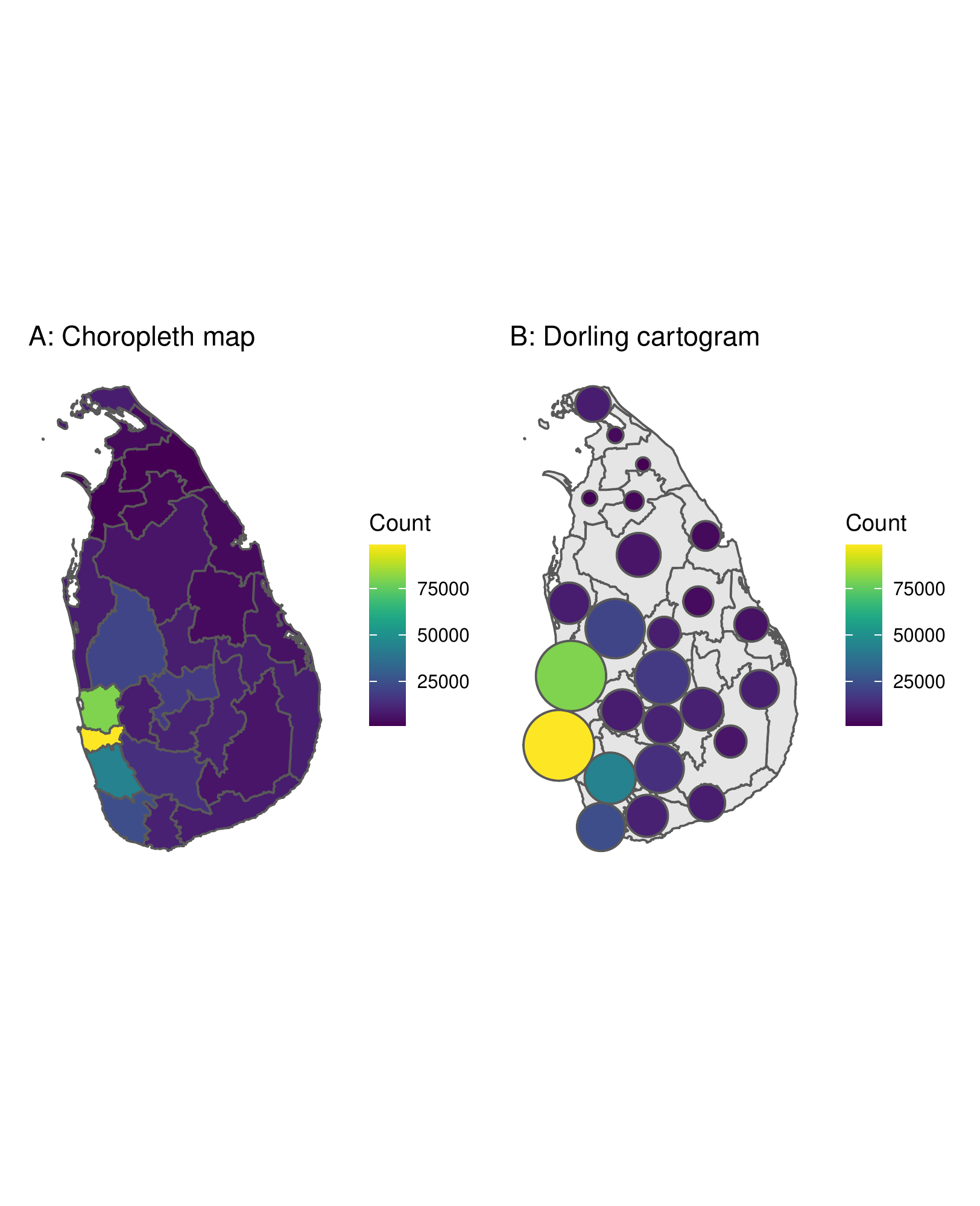}
\caption{Spatial distribution of COVID1-19 cases by districts}
\end{figure}

\hypertarget{results}{%
\section{Results}\label{results}}

The ``Sri Lanka COVID-19 Dashboard'' provides an overview of the
COVID-19 pandemic and administration of vaccine information in Sri
Lanka. This dashboard has eight panels as listed in Table 6.

\textbf{Table 6: Description of the panels}

\begin{longtable}[]{@{}
  >{\raggedright\arraybackslash}p{(\columnwidth - 4\tabcolsep) * \real{0.12}}
  >{\raggedright\arraybackslash}p{(\columnwidth - 4\tabcolsep) * \real{0.80}}
  >{\raggedright\arraybackslash}p{(\columnwidth - 4\tabcolsep) * \real{0.08}}@{}}
\toprule
\textbf{Name of the Panel} & \textbf{Description of the Panel} &
\textbf{Figure} \\
\midrule
\endhead
Overview & Total count of COVID-19 confirmed, recovered, deaths, active
cases and total vaccine doses administered.\hfill\break Provide an
overview of daily COVID-19 confirmed, recovered \& deaths by
plots.\hfill\break & Figure 4 \\
Cases by Wave & There are three tabs in this panel.\hfill\break  * Total
COVID-19 confirmed cases - Cumulative count of COVID-19 confirmed cases
with specific dates\hfill\break * COVID-19 Cases Distribution by Wave -
Daily confirmed cases by wave\hfill\break * COVID-19 Deaths Distribution
by Wave - Daily deaths by wave\hfill\break & Figure 5 \\
COVID-19 Patients Distribution & Provide an overview of confirmed cases
district wise distribution. There are four tabs in this
panel.\hfill\break * Total COVID-19 Patients Distribution in Sri Lanka -
Total confirmed counts for each district is represented by tree
map\hfill\break * Country Map - Total confirmed cases in each district
represented by Sri Lanka country map\hfill\break * Distribution of Daily
COVID-19 Patients for Last 30 Days - Visualize the daily confirmed cases
distribution by districts in last 30 days\hfill\break * By Applying
Min-Max Transformation - Visualize the details in the third tab by
applying min-max transformation for each district\hfill\break & Figure
6 \\
Vaccination Details & Provide an overview of COVID-19 vaccination in Sri
Lanka. There are two tabs.\hfill\break * Total Vaccine Doses - Visualize
the administered vaccine doses as first dose only \& fully
vaccinated\hfill\break * Total Administered Doses by Vaccine Name -
Visualize the vaccination by vaccine names\hfill\break & Figure 7 \\
Top 10 Countries & In this panel, compare the Sri Lanka confirmed \&
deaths with top 10 countries in the world (top 10 countries - The
countries which have been reported highest number of confirmed cases as
31st of August 2021).\hfill\break There are two
tabs.\hfill\break * Comparison of Cumulative Cases in Sri Lanka with Top
10 Countries - Compare the confirmed and deaths in Sri Lanka with top 10
countries by cumulative time series plots\hfill\break * Comparison of
Log of Cumulative Cases in Sri Lanka with Top 10 Countries - Compare the
confirmed and deaths in Sri Lanka with top 10 countries by log
cumulative time series plots\hfill\break (The data has been pulled from
WHO)\hfill\break & Figure 8 \\
Global Comparison & Compare the total confirmed \& deaths in Sri Lanka
with top 10 countries in Global \& Asia. There are two
tabs.\hfill\break * Comparison of the Sri Lanka with Top 10 Countries
Reporting the Most COVID-19 Cases in the World - Compare the total
confirmed \& deaths in Sri Lanka with top 10 countries in the world \&
compare the case fatality ratios\hfill\break * Comparison of the Sri
Lanka with Top 10 Countries Reporting the Most COVID-19 Cases in the
Asia - Compare the total confirmed \& deaths in Sri Lanka with top 10
countries in the Asia \& compare the case fatality ratios\hfill\break &
Figure 9 \\
Global Map & Visualize the distribution of confirmed, recovered \&
deaths in the world by world map.\hfill\break & Figure 10 \\
About & This panel contains the details about the dashboard. & Figure
11 \\
\bottomrule
\end{longtable}

\begin{figure}

{\centering \includegraphics[width=0.8\linewidth]{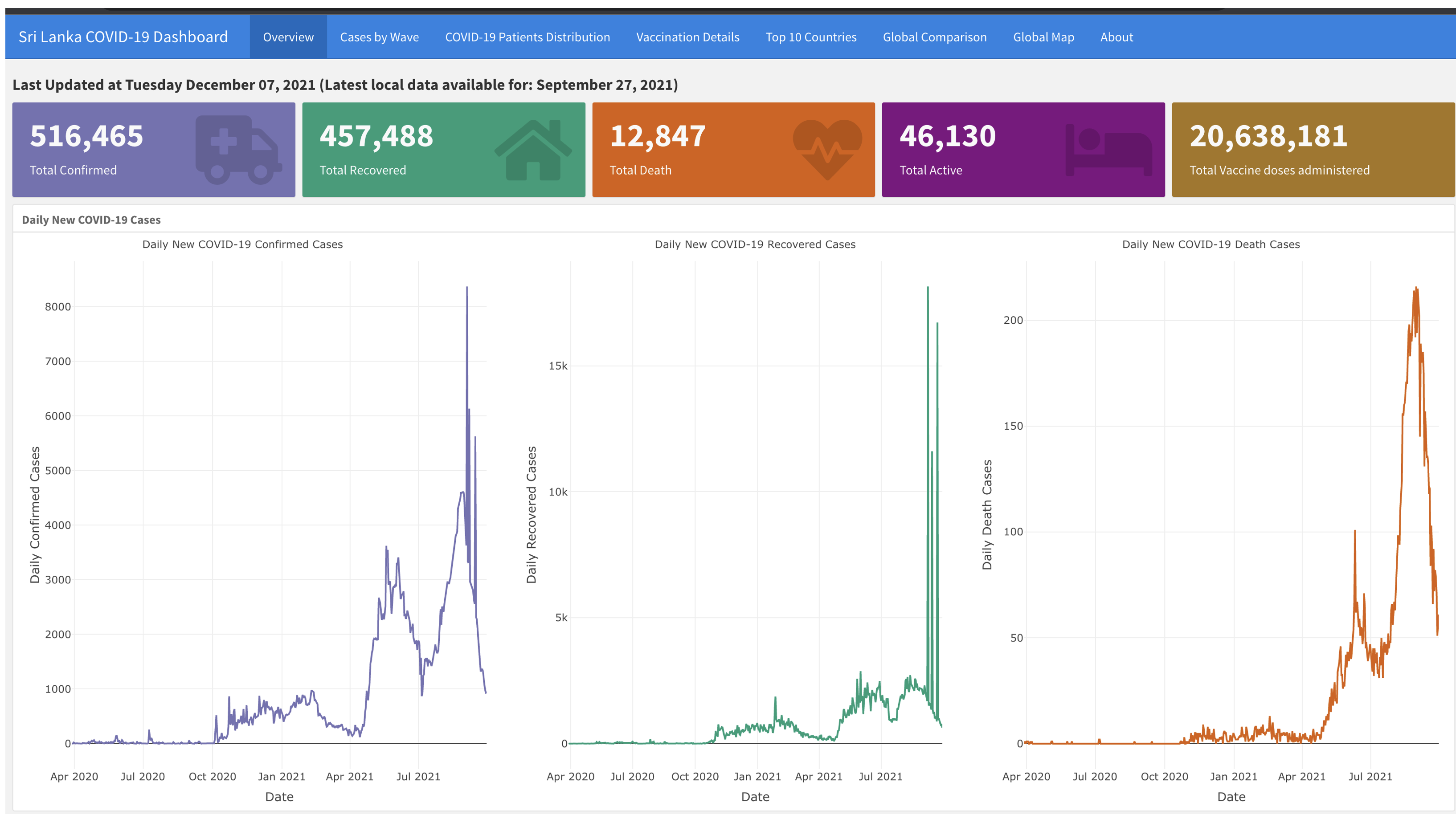} 

}

\caption{Screeshot of panel 1: Overview}\label{fig:unnamed-chunk-5}
\end{figure}

\begin{figure}

{\centering \includegraphics[width=0.8\linewidth]{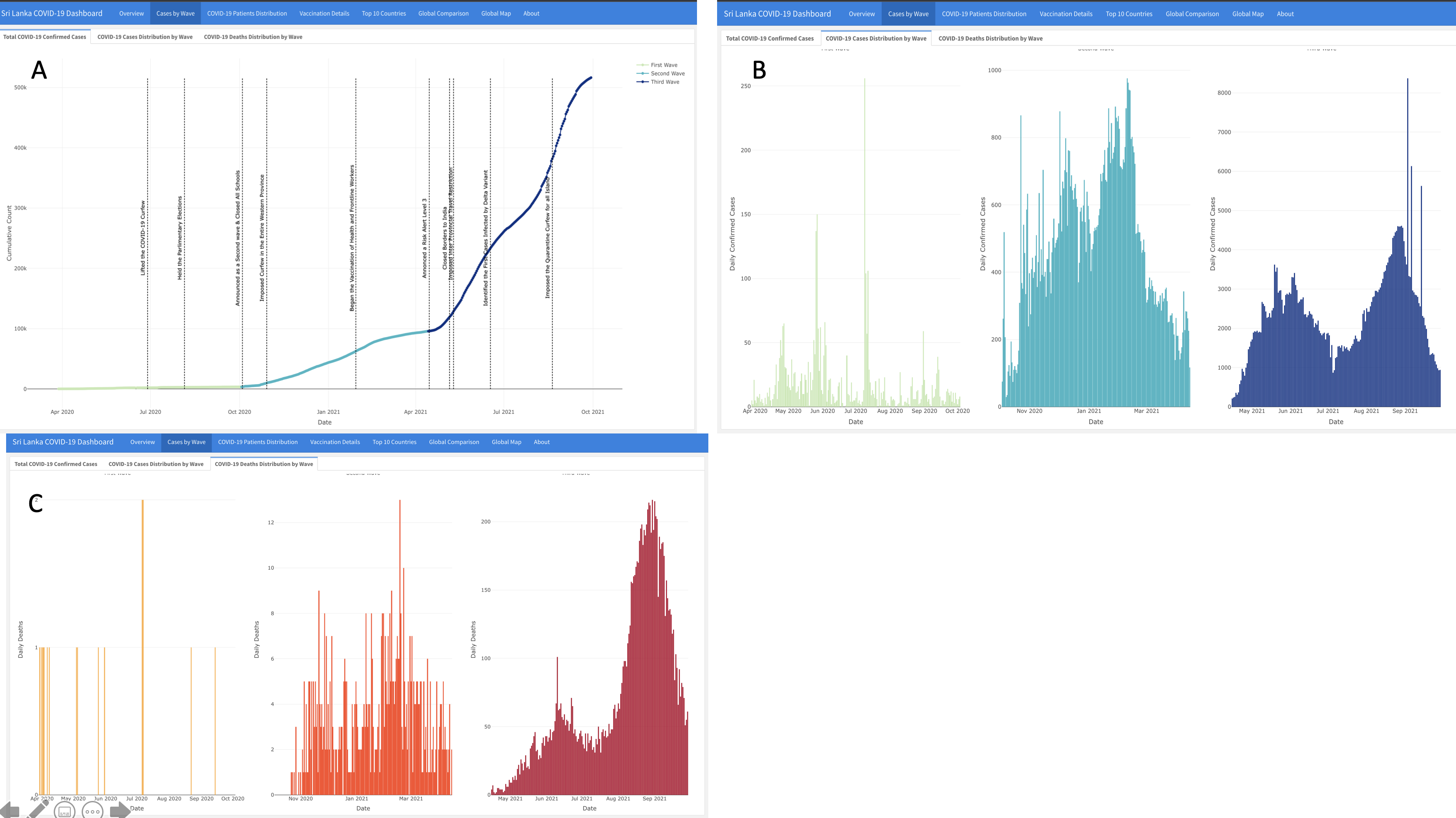} 

}

\caption{Screeshot of panel 2: Cases by Wave}\label{fig:unnamed-chunk-6}
\end{figure}

\begin{figure}

{\centering \includegraphics[width=0.8\linewidth]{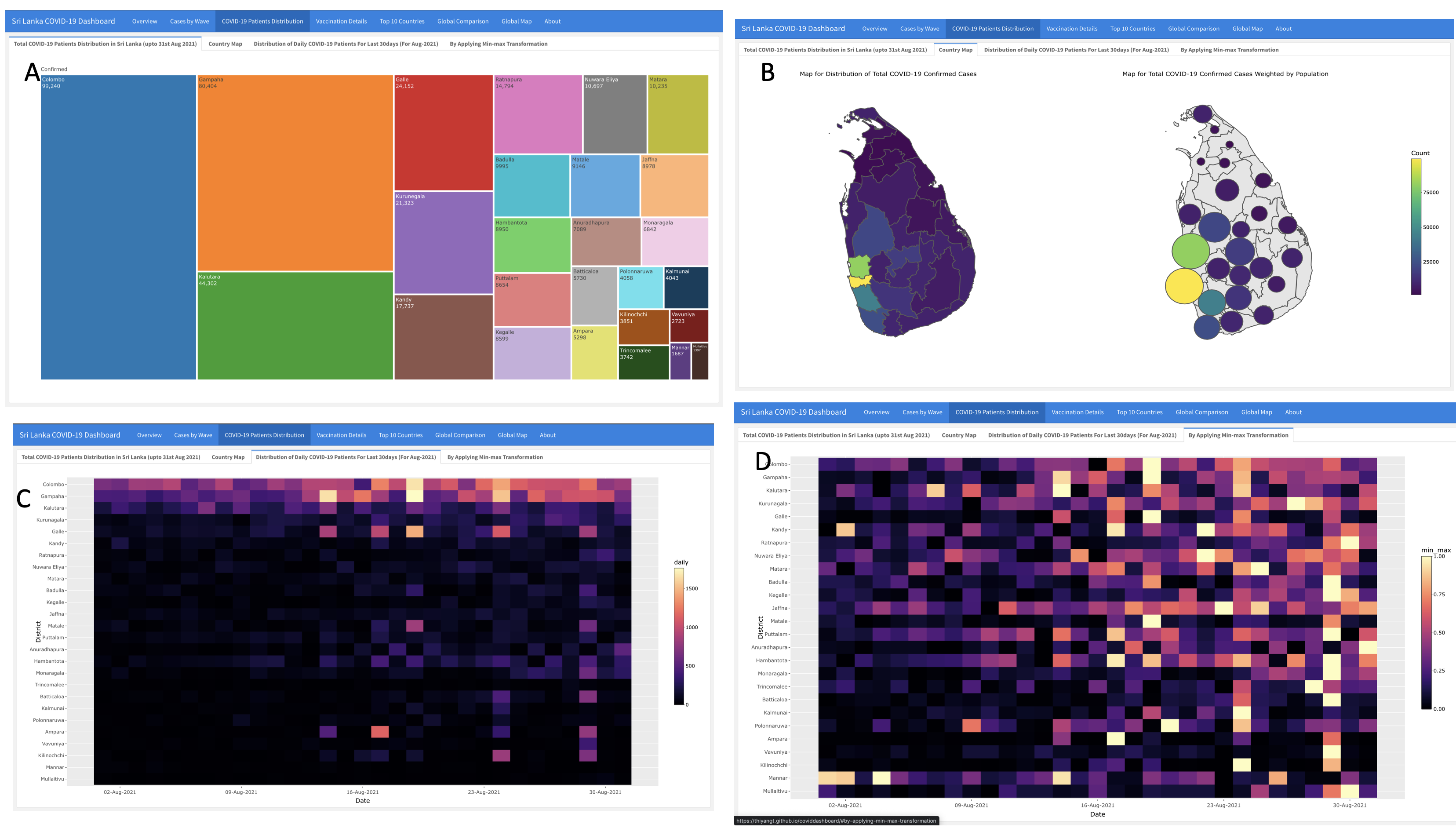} 

}

\caption{Screeshot of panel 3: Distribution of COVID-19 Patients}\label{fig:unnamed-chunk-7}
\end{figure}

\begin{figure}

{\centering \includegraphics[width=0.8\linewidth]{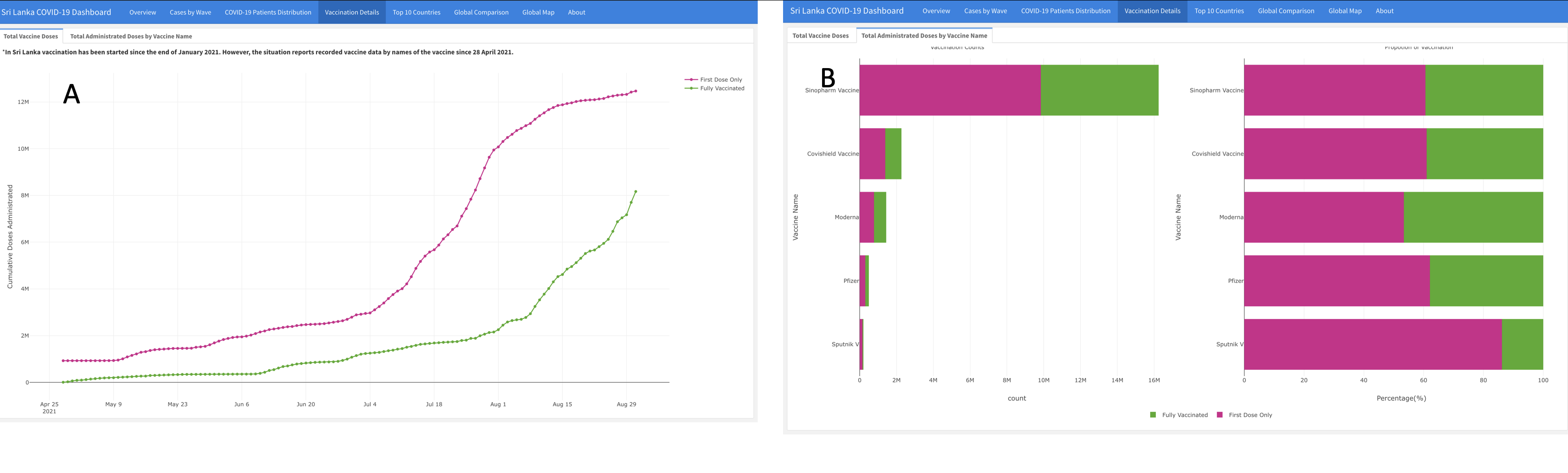} 

}

\caption{Screeshot of panel 4: Vaccination Details}\label{fig:unnamed-chunk-8}
\end{figure}

\begin{figure}

{\centering \includegraphics[width=0.8\linewidth]{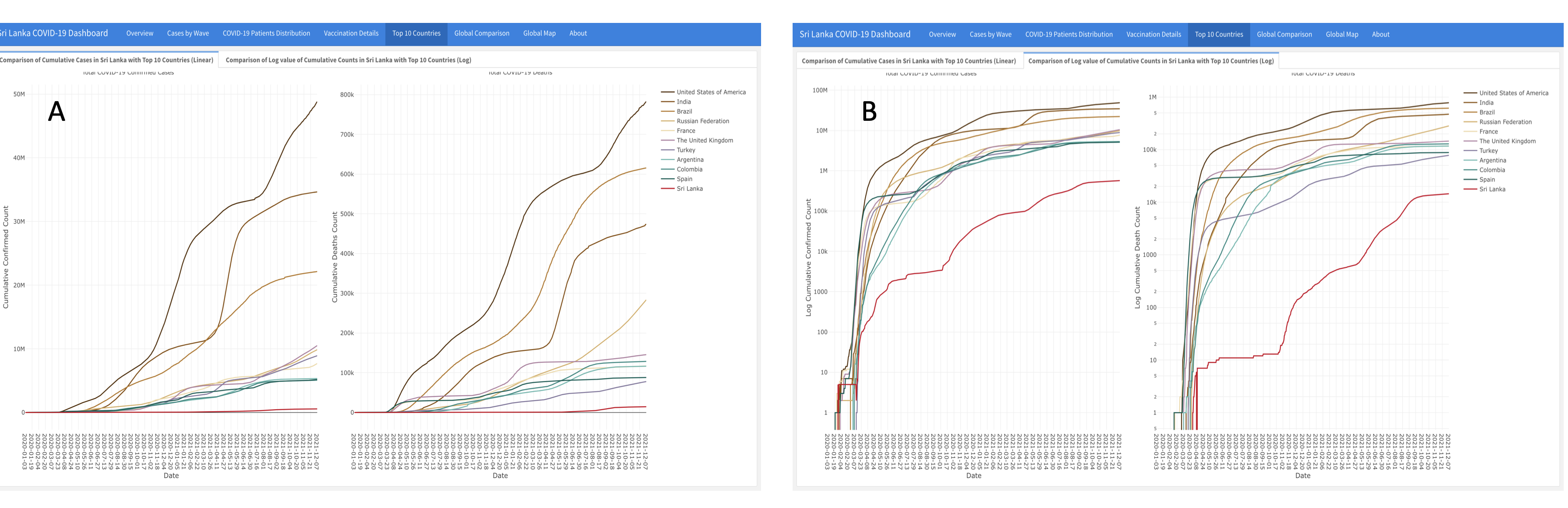} 

}

\caption{Screeshot of panel 5: Comparison with Top 10 Countries}\label{fig:unnamed-chunk-9}
\end{figure}

\begin{figure}

{\centering \includegraphics[width=0.8\linewidth]{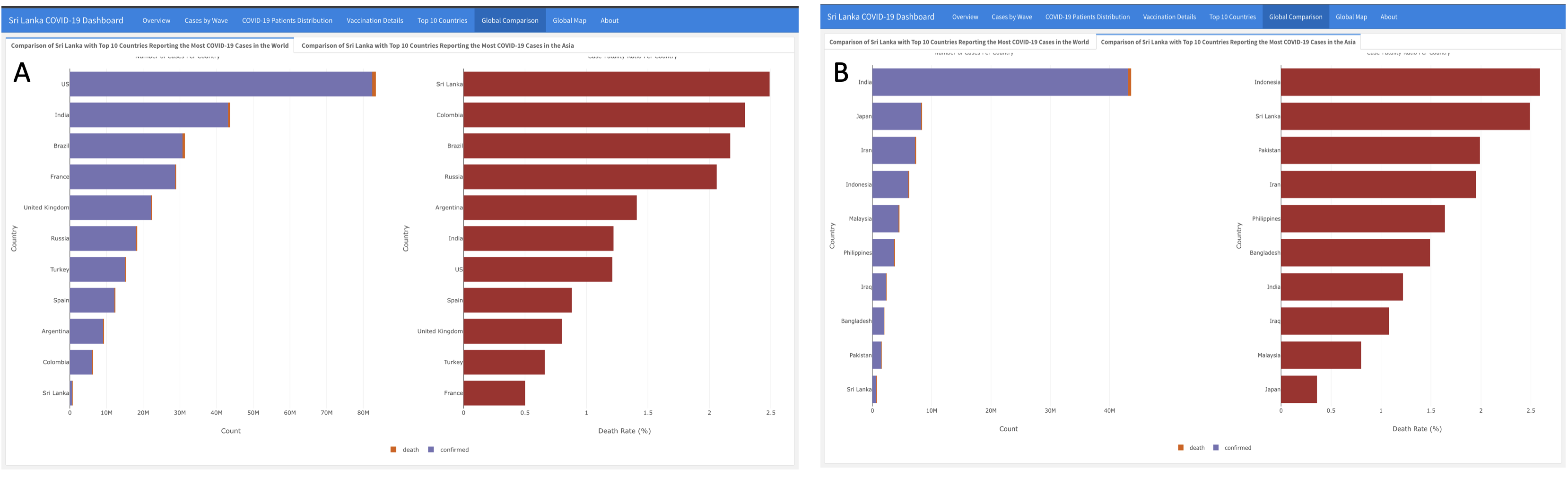} 

}

\caption{Screeshot of Panel 6: Global comparison}\label{fig:unnamed-chunk-10}
\end{figure}

\begin{figure}

{\centering \includegraphics[width=0.8\linewidth]{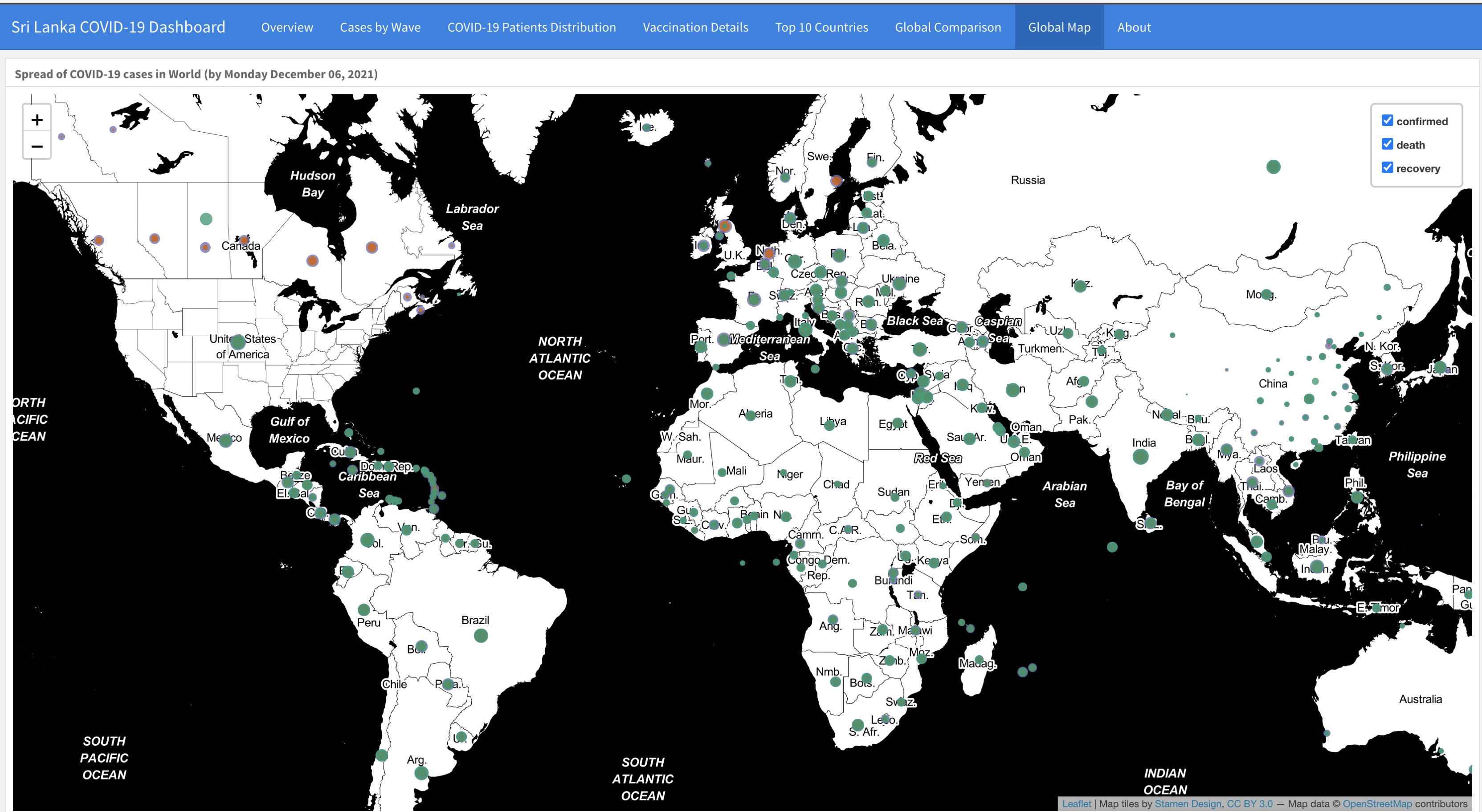} 

}

\caption{Screeshot of panel 7: World Map}\label{fig:unnamed-chunk-11}
\end{figure}

\begin{figure}

{\centering \includegraphics[width=0.8\linewidth]{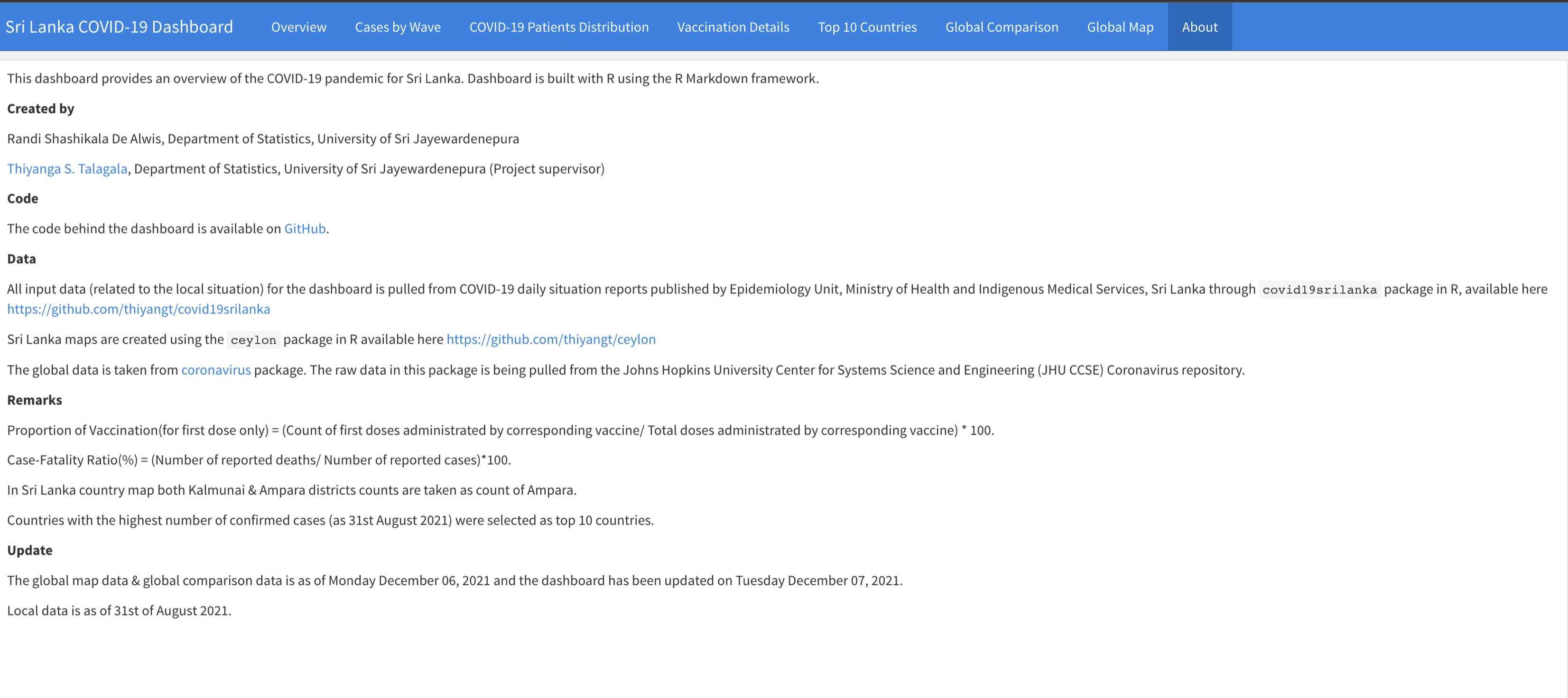} 

}

\caption{Screeshot of panel 8: About}\label{fig:unnamed-chunk-12}
\end{figure}

\newpage

\hypertarget{conclusions}{%
\section{Discussion and Further research}\label{conclusions}}

Bar charts and line charts are the most frequently used tools for the
visualization of total cases, daily cases, and weekly cases, and
comparisons with respect to time. Some dashboards contained
doughnut-shaped pie charts to summarize the total figures. In almost
every dashboard, value boxes have been used to represent total figures.
Some dashboards contained interactive maps and data tables to visualize
the distribution of cases by country, province, region, or state. All
dashboards are updated daily in real time. Gender, age groups, and
ethnicity can be identified as common breakdowns. Most data sets and
related links are available.

We used a colour-blind-friendly theme when creating our dashboard. The
dashboard includes both static and interactive charts that can be used
for explanatory (telling you something), exploratory (give user the
chance to interact with the plots), and exhibitory (showing data
visually) purposes. The data speaks to us, but it is not always easy to
understand in what language they are doing it. Hence, we examine the
same data under different angles using different types of plots. We do
not have district-wise vaccination details. This is a potential future
direction to think about. Our dashboard is completely reproducible. The
source code for reproducing the results is accessible in a public GitHub
repository at \url{https://github.com/thiyangt/covid19srilanka}.

\hypertarget{references}{%
\section{References}\label{references}}

\hypertarget{refs}{}
\begin{CSLReferences}{1}{0}
\leavevmode\hypertarget{ref-belgium}{}%
{``{COVID 19 Dashboard -- Belgium}.''} 2022.
\url{https://esribelux.maps.arcgis.com/apps/ops\%0Adashboard/index.html\#/e350724c87af49bb9ce29646f8a42344}.

\leavevmode\hypertarget{ref-indiadash}{}%
{``{COVID 19 Dashboard India -- ZOHO Analytics -- ZOHO}.''} 2022.
\url{https://www.zoho.com/covid/india/}.

\leavevmode\hypertarget{ref-nhs}{}%
{``{COVID 19 Dashboard -- NHS providers}.''} 2022.
\url{https://nhsproviders.org/topics/covid-\%0A19/tracking-covid-19-data/covid-19-dashboard}.

\leavevmode\hypertarget{ref-italy}{}%
{``{COVID-19 integrated surveillance data in Italy -- EpiCentro}.''}
2022.
\url{https://www.ep\%0Aicentro.iss.it/en/coronavirus/sars-cov-2-dashboard}.

\leavevmode\hypertarget{ref-jamaica}{}%
{``{COVID-19 Jamaica - Ministry of Health and Wellness}.''} 2021.
\url{https://jamcovid19.moh.gov.jm/}.

\leavevmode\hypertarget{ref-sl}{}%
{``{COVID-19: Live situational Analysis Dashboard of Sri Lanka}.''}
2022. \url{https:\%0A//hpb.health.gov.lk/covid19-dashboard/}.

\leavevmode\hypertarget{ref-jh}{}%
{``{COVID-19 Map -- Johns Hopkins Coronavirus Resource Center}.''} 2022.
\url{https://coronavi\%0Arus.jhu.edu/map.html}.

\leavevmode\hypertarget{ref-nssac}{}%
{``{COVID-19 Surveillance Dashboard -- NSSAC Research}.''} 2022.
\url{https://nssac.bii.virg\%0Ainia.edu/covid-19/dashboard/}.

\leavevmode\hypertarget{ref-za}{}%
{``{COVID-19 ZA Dashboard - Data Studio}.''} 2021.
\url{https://datastudio.google.com/u/0/\%0Areporting/1b60bdc7-bec7-44c9-ba29-be0e043d8534/page/hrUIB}.

\leavevmode\hypertarget{ref-flexdashboard}{}%
Iannone, Richard, JJ Allaire, and Barbara Borges. 2020.
\emph{Flexdashboard: R Markdown Format for Flexible Dashboards}.
\url{https://CRAN.R-project.org/package=flexdashboard}.

\leavevmode\hypertarget{ref-rami}{}%
Krispin, Rami. 2021. {``{The Coronavirus Dashboard}.''}
\url{https://github.com/RamiKrispin/coronavirus_dashboard}.

\leavevmode\hypertarget{ref-nz}{}%
{``{New Zealand COVID-19 Surveillance Dashboard}.''} 2021.
\url{https://nzcoviddashboard.esr.c\%0Ari.nz/\#!/}.

\leavevmode\hypertarget{ref-pakistan}{}%
{``{Pakistan's Official COVID-19 Dashboard -- Shifa International
Hospitals Ltd}.''} 2021.
\url{https://www.shifa.com.pk/covid-19-pakistan/}.

\leavevmode\hypertarget{ref-peng2008method}{}%
Peng, Roger D. 2008. {``A Method for Visualizing Multivariate Time
Series Data.''}

\leavevmode\hypertarget{ref-rki}{}%
{``{RKI COVID-19 Germany -- ArcGIS Experience}.''} 2021.
\url{https://experience.arcgis.co\%0Am/experience/478220a4c454480e823b17327b2bf1d4}.

\leavevmode\hypertarget{ref-plotly}{}%
Sievert, Carson. 2020. \emph{Interactive Web-Based Data Visualization
with r, Plotly, and Shiny}. Chapman; Hall/CRC.
\url{https://plotly-r.com}.

\leavevmode\hypertarget{ref-talagala}{}%
Talagala, Thiyanga S. 2021. \emph{Covid19srilanka: The 2019 Novel
Coronavirus COVID-19 (2019-nCoV) Data in Sri Lanka}.
\url{https://github.com/thiyangt/covid19srilanka}.

\leavevmode\hypertarget{ref-russia}{}%
{``{The Global COVId-19 Index (GCI) -- Russia Dashboard -- PEMANDU
Associates}.''} 2021. \url{https://covid19.pemandu.org/Russia}.

\leavevmode\hypertarget{ref-canada}{}%
{``{Track COVID-19 Across Canada Using Our Interactive Dashboards}.''}
2021.
\url{https:\%0A//samples.dundas.com/Dashboard/62cef916-5488-4d3b-9c57-204092d01813?e=false\&vo=viewonly}.

\leavevmode\hypertarget{ref-who}{}%
{``{WHO Coronavirus (COVID-19) Dashboard}.''} 2021.
\url{https://covid19.who.int/region/amro/country/us}.

\end{CSLReferences}


@Book{plotly,
    author = {Carson Sievert},
    title = {Interactive Web-Based Data Visualization with R, plotly, and shiny},
    publisher = {Chapman and Hall/CRC},
    year = {2020},
    isbn = {9781138331457},
    url = {https://plotly-r.com},
  }
\end{document}